\documentclass[10pt, twoside, twocolumn]{article}
\usepackage {psfig}
\usepackage {epsf}

\newcommand {\bc}{\begin {center}}
\newcommand {\ec}{\end {center}}
\newcommand {\be}{\begin {equation}}
\newcommand {\ee}{\end {equation}}
\newcommand {\beq}{\begin {eqnarray}}
\newcommand {\eeq}{\end {eqnarray}}

\newcommand {\sh}{\mathop {\rm sh}\nolimits}
\newcommand {\ch}{\mathop {\rm ch}\nolimits}
\newcommand {\ctg}{\mathop {\rm ctg}\nolimits}
\newcommand {\tg}{\mathop {\rm tg}\nolimits}
\newcommand {\cth}{\mathop {\rm cth}\nolimits}
\newcommand {\aaa}{\mathop {\rm sn}\nolimits}
\newcommand {\rrr}{\mathop {\rm cs}\nolimits}
\newcommand {\uuu}{\mathop {\rm sc}\nolimits}
\newcommand {\www}{\mathop {\rm cn}\nolimits}
\newcommand {\ar}{\mathop {\rm ar}\nolimits}

\newcommand {\arsh}{\mathop {\rm arsh}\nolimits}
\def\st{\star}
\def\disp{\displaystyle}
\def\prt{\partial}
\def\ff{\varphi}
\def\eps{\epsilon}
\def\tet{\vartheta}
\def\intl{\int\limits}

\def\cI {{\cal I}}

\def\b {{\rm b}}
\def\c {{\rm c}}
\def\d {{\rm d}}
\def\e {{\rm e}}
\def\i {{\rm i}}
\def\m {{\rm m}}
\def\o {{\rm o}}
\def\r {{\rm r}}
\def\s {{\rm s}}
\def\uu {{\rm u}}
\def\H {{\rm H}}
\def\T {{\rm T}}
\def\W {{\rm W}}
\def\kB {k_{\rm B}}
\def\DR#1#2{\frac {\d#1}{\d#2}}
\def\Dr#1#2{\frac {\prt#1}{\prt#2}}

\begin {document}
\title{Sources of Radiation in the Early Universe: The Equation \\
of Radiative Transfer and Optical Distances} \date{}  
\twocolumn[
\begin{@twocolumnfalse}
\begin{flushright}
{\it Astronomy Reports, Vol.49, No 3, 2005,\\ pp. 167-178.
Translated from Astronomicheski$\tilde\imath$ \\ Zhurnal, Vol. 82, No 3, 2005,
pp.195-206.\\ Original Russian Text Copiright \\ \copyright 2005 by Nagirner, 
Kirusheva.}
\end{flushright}

\maketitle

\bc
D.~I.Nagirner
S.~L.Kirusheva
\ec
\bc
\it {Sobolev Astronomical Institute, St. Petersburg State
University, Bibliotechnaya pl. 2, Petrodvorets, 198904 Russia}
\ec
\begin {abstract}
We have derived the radiative-transfer equation for a point source with a specified intensity 
and spectrum, originating in the early Universe between the epochs of annihilation and recombination, 
at redshifts $z_\s =10^8\div 10^4$. The direct radiation of the source is separated from the diffuse radiation it 
produces. Optical distances from the source for Thomson scattering and bremsstrahlung absorption at the 
maximum of the thermal background radiation are calculated as a function of the redshift z.The distances 
grow sharply with decreasing z, approaching asymptotic values, the absorption distance increasing more 
slowly and reaching their limiting values at lower z. For the adopted z values, the optical parameters of the 
Universe can be described in a flat model with dusty material and radiation, and radiative transfer can be 
treated in a grey approximation. 
\end {abstract}
 \end{@twocolumnfalse}
]
\pagestyle{myheadings}
\markboth{ NAGIRNER, KIRUSHEVA }{SOURCES OF RADIATION IN THE EARLY UNIVERSE}
\section{Formulation of the problem}
Let the early Universe be described by the standard
 model, i.e. by the Friedmann equations (see, 
for example, \cite{ZelNov}). Let us suppose that a source of 
radiation is switched on at some time between the 
epochs of annihilation and recombination, that this 
source has a specified geometry, spectrum, and total 
luminosity, and that it radiates during some time 
interval. The intensity and spectrum of the source 
may be time-dependent. Such sources could occur in 
a number of processes, for example, during the amplification
 or damping of fluctuations in the matter distribution,
 in the formation of primary black holes, etc. 

The action of the source causes the thermodynamical
 equilibrium in its vicinity to be violated, since the 
source radiation is either nonequilibrium (for example,
 it has a power-law spectrum) or corresponds to 
a higher temperature. The radiation can experience 
bremsstrahlung absorption and Thomson scattering. 
In \cite{dubr}, it was suggested that such sources could affect
 the background radiation and hence be observed 
via perturbations of this background, contrary to the 
prevalent opinion that the energy of such sources is 
totally dissipated. Both possibilities require careful 
verification using the methods of radiative-transfer 
theory. 

We assume that the source is pointlike. The pointsource
 problem is also of independent interest: it is 
fundamental, since all other types of sources can be 
reduced to a set of point sources. 

If the source radiation is intense, it affects the 
surrounding matter in several ways. First, the radiation
 creates an additional pressure, which affects the 
local expansion of space. Second, this same pressure 
gives rise to inhomogeneity of the expansion. Third, 
a shock and, in the case of a periodic source, matter 
oscillations may be formed. Fourth, the source can 
distort the metric of the surrounding space. In all 
these cases, the geometry of the source is the key 
factor. 

Here, we will assume that the source radiation is 
moderately intense and does not affect the expansion
 of the surrounding space. To start with, we will 
present the model for the Universe at the considered 
epoch. 
\section{Two-component model of the Universe: 
dustlike matter and radiation }
\subsection{BasicRelations }
We consider a model with a homogeneous and 
isotropic expanding Universe that contains equilibrium plasma and Planck 
radiation in the stage when the plasma is radiation-dominated; more 
exactly, between the epochs of annihilation ($z<10^8$) and recombination
($z>10^3$). In this stage of the evolution of the Universe, the 
cosmological term does not play a marked role. 

At the temperatures corresponding to these redshifts, the matter can 
be assumed to be nonrelativistic, and the conditions of thermodynamical 
balance with a common temperature for the matter and radiation are 
fulfilled \cite{ZelNov}.  The pressure of the matter may be neglected, 
i.e., the matter can be considered to be dustlike, in spite of its nearly 
total ionization.

Under these conditions, in the usual notation, 
the evolution of the expansion is described by the 
equation 

\begin {equation}
\dot {R}^2=\frac {8\pi G}{3}\rho_\uu R^2-kc^2,
\label{1}
\end {equation}
where a dot denotes a derivative with respect to the
time t, the total mass density of the matter is equal
to the sum of the matter and radiation densities, $\rho_\uu=\rho_\d+\rho_\r$,
and the parameter $k$ assumes the values 
\begin{eqnarray*}
k=\left\{
\begin {array}{lll}
1, & \hbox {\rm for a closed model}, \\
0, & \hbox {\rm for a flat model}, \\
-1, & \hbox {\rm for an open model}. \\
\end {array}\right.
\end{eqnarray*}
The matter density decreases in inverse proportion to the volume, 
while the expression for the radiation density contains another 
power of the radius of curvature in the denominator, resulting 
from the cosmological  redshift: 
\begin {eqnarray*} 
\rho_\d=\rho_\d^0\frac {R_0^3}{R^3}=\frac {\rho_\d^0}{a^3},\quad
\rho_\r=\rho_\r^0\frac {R_0^4}{R^4}=\frac {\rho_\r^0}{a^4}.
\end {eqnarray*}
Here and below, zero superscripts will denote cosmological values 
at the current epoch. 

We will use the dimensionless time coordinate 
$\eta$ and dimensionless radius of curvature (scale factor) $a(\eta)$:
\begin {equation}
R(t)\d\eta=c\d t,\quad a(\eta)=\frac {R(t)}{R_0}.
\label{2}
\end {equation}
With these variables, Eq. (1) can be transformed into 
\begin {equation}
\left(\DR {a}{\eta}\right)^2=\frac {8\pi G}{3c^2}(\rho_\d^0 a+\rho_\r^0)R_0^2-
ka^2.
\label{3}
\end {equation}
This model was first considered by Chernin \cite{Arthur}. 

Let us also introduce the Hubble function and the 
critical density expressed in terms of this function:
\begin {eqnarray*}
&\disp H=\frac {\dot {R}}{R}=\frac {\dot {a}}{a}=
\frac {c}{a}\frac {1}{R}\DR {a}{\eta}=\frac {c}{R_0}
\frac {a'(\eta)}{a^2(\eta)},& \\
&\disp \rho_\c=\frac {3H^2}{8\pi G}.&
\end {eqnarray*}
Along with the density itself, we will use the critical 
parameters 
\begin {equation}
\Omega_\d=\frac {\rho_\d}{\rho_\c},\quad \Omega_\r=\frac {\rho_\r}{\rho_\c},
\quad \Omega_\uu=\Omega_\d+\Omega_\r,
\label{4}
\end {equation}
which can be used to rewrite (1) in another form: 
\begin {equation}
R^2H^2(1-\Omega_\uu)=-kc^2.
\label{5}
\end {equation}

\subsection{AuxiliaryFunctions }
 Let us define several mutually related functions, in terms of which 
the solutions and relations of the model will be written. The two 
main functions are $\aaa_k(\eta)$ and $\uuu_k(\eta)$, which are defined 
by the equalities 
\begin {equation}
\aaa_k(\eta)=\left\{
\begin {array}{ll}
\sin\eta & \hbox {\rm when} \quad k=1,    \\
\eta     & \hbox {\rm when} \quad  k=0,    \\
\sh\eta  & \hbox {\rm when} \quad  k=-1,   \\
\end {array}\right.
\label{6}
\end {equation}

\begin {eqnarray*} 
\uuu_k(\eta)=\left\{
\begin {array}{lcl}
1-\cos\eta & \hbox {\rm when} \quad  k=1, \\
\eta^2/2   & \hbox {\rm when} \quad  k=0, \\
\ch\eta-1  & \hbox {\rm when} \quad  k=-1, \\
\end {array}\right. 
\end {eqnarray*}
\begin {eqnarray*}
\uuu_k'(\eta)=\aaa_k(\eta).
\end {eqnarray*}
Two functions are related to function (6). One of 
these, 
\begin {eqnarray}
\rrr_k(\eta)=\aaa'_k(\eta)=\left\{
\begin {array}{lll}
\cos\eta & \hbox {\rm when} & k=1,    \\
1        & \hbox {\rm when} & k=0,    \\
\ch\eta  & \hbox {\rm when} & k=-1,   \\
\end {array}\right.
\label{7}
\end {eqnarray}
is its derivative, while the other, 
\begin {eqnarray*}
\ar_k(y)\!=\!\left\{
\begin {array}{ll}
\!\arcsin y, & k=1, \\
\! y, & k=0, \\
\!\arsh y=\ln(y+\sqrt {1+y^2}), & k\!=\! -1,
\end {array}\right.
\end {eqnarray*}
\begin {eqnarray*}
\disp \DR {\ar_k(y)}{y}=\frac {1}{\sqrt {1-ky^2}}
\end {eqnarray*}
is the inverse of $\aaa_k(\chi)$, i.e.,
$\ar_k(\aaa_k(\eta))=\eta,\,\,\aaa_k(\ar_k(y)))=y$.

Another function,
\[
\www_k(\eta)=\left\{
\begin {array}{ll}
\eta-\sin\eta & \hbox {\rm when}\quad  k=1,    \\
\eta^3/6      & \hbox {\rm when}\quad   k=0,    \\
\sh\eta-\eta  & \hbox {\rm when}\quad   k=-1,   \\
\end {array}\right.
\]
is the integral of $\uuu_k(\eta)$,i.e., $\www'_k(\eta)=\uuu_k(\eta)$. 

 It can easily be verified that the following relations 
between the introduced functions are satisfied: 
\begin {eqnarray}
\rrr_k^2+k\aaa_k^2=1, & \label{8}  \\
\aaa_k^2+k\uuu_k^2=2\uuu_k, &\rrr_k+k\uuu_k=1.\nonumber
\end {eqnarray}
Let us consider the ratio 
\begin {eqnarray}
y(\eta)=\frac {\aaa_k(\eta)}{\uuu_k(\eta)}=\left\{
\begin {array}{llll}
\disp\ctg\frac {\eta}{2} & \hbox {\rm when} & k=1, \\
\disp\frac {2}{\eta}     & \hbox {\rm when} & k=0, \\
\disp\cth\frac {\eta}{2}  & \hbox {\rm when}& k=-1. \\
\end {array}\right.
\label{9}
\end {eqnarray}
Its derivative with respect to $\eta$ assumes a simple form 
if we use the relation between the derivatives and the
functions and equalities (\ref{8}):
\begin {eqnarray}
y'(\eta)\! =\!\frac {\uuu_k-k\uuu^2_k-2\uuu_k+k\uuu^2_k}{\uuu^2_k}=\!-
\frac {1}{\uuu_k}.
\label{10}
\end {eqnarray}
Using the second of the equalities, this ratio can, in 
turn, be expressed in terms of y: 
\begin {equation}
\DR {y}{\eta}=-\frac {y^2+k}{2}.
\label{11}
\end {equation}
To obtain the correct solution, we must apply the 
appropriate initial condition: $y=\infty$ when $\eta=0$.The 
arguments of the functions in relations (8), (10), 
and (11) have been omitted for brevity. 
\subsection{Solution of the Equation of Motion }
It was shown in \cite{Arthur} that the solution of (3) in 
implicit form is 
\begin {eqnarray}
&\disp a(\eta)=a_\r\aaa_k(\eta)+a_\d\uuu_k(\eta), & \nonumber\\
&\disp t=\frac {R_0}{c}[a_\r\uuu_k(\eta)+a_\d\www_k(\eta)]. &
\label{12}
\end {eqnarray}
Indeed, we can easily verify using relations (8) that 
the function  $a(\eta)$ satisfies (3) if we assume 
\[
a_\r=\sqrt {\frac {8\pi G}{3}\rho_\r^0}\frac {R_0}{c},\quad
a_\d=\frac {4\pi G}{3c^2}\rho_\d^0R_0^2.
\]
The obtained coefficients are expressed in terms of the 
critical parameters. Let us first derive an expression
for the current radius of curvature \cite{ZelNov}. To this end, we 
apply formula (5) to the current epoch: 
\begin{eqnarray*}
H_0^2R_0^2=-\frac {kc^2}{1-\Omega_\uu^0},\quad \frac {H_0R_0}{c}=
\frac {1}{\sqrt {|1-\Omega_\uu^0|}}.
\end{eqnarray*}
We now have, in accordance with the definitions (4), 
\begin {eqnarray*}
&\disp a_\d=\frac {4\pi G}{3c^2}\rho_\c^0\Omega_\d^0R_0^2=\frac {4\pi G}{3c^2}
\frac {3H_0^2}{8\pi G}\Omega_\d^0R_0^2 & \\
&\disp =-\frac {k}{2}\frac {\Omega_\d^0}
{1-\Omega_\uu^0}=\frac {\Omega_\d^0}{2|1-\Omega_\uu^0|}, & \\
&\disp a_\r^2=\frac {8\pi G}{3c^2}\rho_\c^0\Omega_\r^0R_0^2=\frac {8\pi G}{3c^2}
\frac {3H_0^2}{8\pi G}\Omega_\d^0R_0^2 & \\
&\disp =-\frac {k\Omega_\r^0}
{1-\Omega_\uu^0}=\frac {\Omega_\r^0}{|1-\Omega_\uu^0|}.&
\end {eqnarray*}
The inverse formulas have the form: 
\begin {eqnarray*}
&\disp \Omega_\d^0=\frac {2a_\d}{2a_\d+a_\r^2-k},\quad
\Omega_\r^0=\frac {a_\r^2}{2a_\d+a_\r^2-k}, & \\
&\disp \Omega_\uu^0=\frac {2a_\d+a_\r^2}{2a_\d+a_\r^2-k}.&
\end {eqnarray*}
In the following subsections (to Subsection 9), we will assume 
that $k\neq 0$, although some functions are calculated using 
formulas that include the special case of a flat space. 

\subsection{Relation to the Redshift}
Let us derive expressions for the quantities associated
 with the solution (\ref{12}) in terms of the redshift $z$. By definition 
$\disp a(\eta)=\frac {R(\eta)}{R_0}=\frac {1}{1+z}$.
Successively substituting in (\ref{12}) the expression for $\aaa_k$ or  $\uuu_k$ 
in terms of the other, and also using the last relation in  (\ref{8}), 
we obtain 
\begin {eqnarray*}
&\disp \aaa_k(\eta)=\frac {Q-B(1-kA)}{1+kB^2}, & \\
&\disp \uuu_k(\eta)=\frac {B^2+A-BQ}{1+kB^2},& \\
&\disp \rrr_k(\eta)=\frac {1-kA+kBQ}{1+kB^2}. &
\end {eqnarray*}
Here, we have introduced the notation 
\begin {eqnarray}
&\disp A=\frac {1}{a_\d}\frac {1}{1+z}=\frac {2|1-\Omega_\uu^0|}{\Omega_\d^0}
\frac {1}{1+z}, \nonumber &\\
&\disp B=\frac {a_\r}{a_\d}=\frac {2\sqrt
{\Omega_\r^0|1-\Omega_\uu^0|}}{\Omega_\d^0},&\\
&\disp Q=\sqrt {B^2+2A-kA^2}. \nonumber &
\label{13}
\end {eqnarray}
The derivative $a'(\eta)$  and the Hubble function can be 
 expressed in terms of the redshift in this same way: 
\begin {eqnarray*}
&\disp a'(\eta)=a_\r\rrr_k(\eta)+a_\d\uuu_k(\eta), &\\
&\disp \frac {H}{H_0}=\frac {c}{H_0R_0}\frac {a'(\eta)}{a^2(\eta)}=
\frac {a'(\eta)}{a^2(\eta)}\sqrt {|1-\Omega_\uu^0|}.&
\end {eqnarray*}
The ratio (\ref{9}) and the product of this ratio with $B$
can also be expressed in terms of $z$: 
\begin {eqnarray*}
y(\eta)=\frac {kAB-B+Q}{B^2+A-BQ},\\
X=By=\frac {kAB-B+Q}{B+A/B-Q}.
\end {eqnarray*}
The critical parameters at the current epoch ($B$ is constant) can 
likewise be expressed in terms of the quantities in (\ref{13}):
\begin {eqnarray*}
&\disp \Omega_\d^0=\frac {2A_0}{Q_0^2},\quad \Omega_\r^0=\frac {B^2}{Q_0^2},&\\
&\disp \Omega_\uu^0=\frac {2A_0+B^2}{Q_0^2},&\\
&\disp 1-\Omega_\uu^0=-k\frac {A_0^2}{Q_0^2}.&
\end {eqnarray*}
It is easy to verify that the identities  $a(\eta_0)=1$ and
$a'(\eta_0)=H_0R_0/c=1/\sqrt{|1-\Omega_\uu^0|}$ are satisfied. 

\subsection{SpecificModels }
According to the above model, the curvature of 
space can be arbitrary (constant). However, recent 
observations lead to the conclusion that space was 
flat over the entire history of the Universe. In this 
context, we will consider two specific models. One is 
open, close to flat, and does not take into account the 
vacuum component (we will call this the nonvacuum 
model), while the other has three components and 
is strictly flat, but with the vacuum neglected at $z>10^3$.

We will adopt for the Hubble constant $H_0=65$~km\,s$^{-1}$\,Mpc$^{-1}$,
the corresponding critical density 
$\rho_\c^0=7.940\times 10^{-30}$~{g/cm$^3$}, and the critical 
parameters  for the baryon component $\Omega_\b^0=0.025$ and dustlike
 component $\Omega_\d^0=0.25$. The baryon density is then 
$\rho_\b^0=4.684\times 10^{-31}$~{g/cm$^3$}, the density of dustlike
matter is $\rho_\d^0=4.684\times 10^{-30}$~{ g/cm$^3$}, the mass 
density of the radiation corresponding to the temperature
 $T_0=2.7277$~K, is $\rho_\r^0=4.63\times 10^{-34}$~{g/cm$^3$}, and 
$\Omega_\r^0=5.83\times 10^{-5}$.
\subsubsection{ Nonvacuum model.} For the adopted model 
parameters,   $1-\Omega_\uu^0=1-\Omega_\d^0-\Omega_\r^0=0.7499$; i.e., this 
model is open ($k=-1$), which is natural, since there is no vacuum component. 
The other parameters are  $a_\r=8.869\times 10^{-3}$ and  $a_\d=0.1667$.

\subsubsection{Strictly flat model.} For $k=0$ ,we also have 
$1-\Omega_\uu=0$, so that many of the introduced values 
become equal to zero or infinity. In this case, the limit 
transition is complicated, and it is easier to consider 
this case individually. 

We will assume that the most plausible model of the Universe for 
redshifts less than 10--100, in particular for the current epoch, 
is a flat model with three noninteracting components: dustlike 
matter, radiation, and the vacuum. Based on this model, we will 
also construct a two-component flat model for the considered epoch 
of radiation-dominated plasma in which the two components have the 
same densities, but there is no vacuum component. All values corresponding
to this model will be denoted by a tilda. The contribution of the vacuum 
can be neglected for the redshifts considered ($10^3\!\leq\!z\!\leq\!10^8$), 
so that the Hubble functions in the two-component and 
three-component flat models differ only slightly: 
\begin {eqnarray*}
&\disp H=\sqrt {\frac {8\pi G}{3}}\sqrt {\rho_\d^0(1+z)^3+\rho_\r^0(1+z)^4+
\rho_\Lambda},&\\
&\disp \tilde {H}=\sqrt {\frac {8\pi G}{3}}\sqrt
{\rho_\d^0(1+z)^3+\rho_\r^0(1+z)^4}.&
\end {eqnarray*}
We will use the same critical parameters $\Omega_\d^0$ and $\Omega_\r^0$
as above; the critical parameter for the vacuum is then 
$\Omega_\Lambda^0=1-\Omega_\d^0-\Omega_\r^0$. In the adopted model, these 
values are: 
\begin {eqnarray}
\tilde {H}_0=H_0\sqrt {\frac {\rho_\d^0+\rho_\r^0}{\rho_\d^0+\rho_\r^0+
\rho_\Lambda^0}}=H_0\sqrt {1-\Omega_\Lambda^0},\label{141} \\
\tilde{\rho}_\c^0=\rho_\c^0(1-\Omega_\Lambda^0), \label{142}\\
\tilde {\Omega}_\d^0=\frac {\Omega_\d^0}{1-\Omega_\Lambda^0},
\tilde {\Omega}_\r^0=\frac {\Omega_\r^0}{1-\Omega_\Lambda^0}.\label{143}
\end {eqnarray}
For brevity, let us set $\tilde{\Omega}_\d=\Omega$, in which case 
$\tilde{\Omega}_\r=1-\Omega$, so that $\tilde{\Omega}_\d^0=\Omega_0$
and $\tilde{\Omega}_\r^0=1-\Omega_0$.

The radius of curvature $R$ and its current value $R_0$
are meaningless in a flat Universe (they are infinite, 
and can be selected arbitrarily in the formulas), and 
should not arise in expressions for physical values. 
Indeed, the solution (\ref{12}) can be written in the form 
\begin {eqnarray*}
&\disp a(\eta)=a_\r\eta+a_\d\frac {\eta^2}{2}=\sqrt {1-\Omega_0}
\frac {\tilde {H}_0R_0}{c}\eta &\\
&\disp +\Omega_0\frac {\tilde {H}_0^2R_0^2}{4c^2}\eta^2
=2\sqrt {1-\Omega_0}\zeta+\Omega_0\zeta^2,&
\end {eqnarray*}
where we have introduced a new time variable that 
is linearly related to the previous time variable and is 
expressed in terms of $a(\eta)$ and the redshift: 
\begin {eqnarray*}
&\disp \zeta\!=\!\frac {\tilde {H}_0R_0}{2c}\eta=\frac {a}{\sqrt
{1-\Omega_0(1-a)}+
\sqrt {1-\Omega_0}} &\\
&\disp \!=\!\!\frac {1}{\!\sqrt {1\!+\!z}\!\left(\!\sqrt {1\!+\!(1\!-\!\Omega_0)z}\!+\!
\sqrt {(1\!-\!\Omega_0)(1\!+\!z)}\right)}.&
\end {eqnarray*}
\section{ Radiative-transfer equation in the early Universe}
\subsection{ General Form of the Radiative-Transfer
Equation for a Point Source}
Let us assume the source to be pointlike and isotropic. In this case,
 the problem is spherically symmetrical, so that, generally speaking, 
the approach of the Tolman-Bondi model should be used. However, to first 
approximation, we can assume that the source radiation does not affect 
the metric of space and its expansion, and the evolution of the source 
radiation can be considered against the background of the standard 
cosmological two-component model described in Section 2.

We will describe the radiation with the mean occupation number of photon 
states n, rather then the intensity. The advantage of this quantity is 
that it is dimensionless and relativistically invariant (scalar). Due 
to the symmetry of our problem, the mean occupation number of photon 
states will depend on the time $t$ or time variable $\eta$, the distance 
of the point considered from the coordinate origin (described by the 
parameter $\chi$), the angle $\theta$ between the radial direction and 
the ray along which the radiation propagates, and the dimensionless 
frequency $x$: $n=n(\eta,\chi,\theta,x)$.
The radiative-transfer equation has the form
\begin {eqnarray}
\!\DR {n}{t}\!=\!\Dr {n}{t}\!+\!\Dr {n}{x}\DR {x}{t}\!+\!\Dr {n}{\chi}\DR
{\chi}{t}\!+\!
\Dr {n}{\theta}\DR {\theta}{t}\!=\! c\cI_\c,
\label{15}
\end {eqnarray}
where $\cI_\c$ is the collision integral, i.e., the difference between 
the numbers of photons with specified parameters entering and leaving 
the state. 

We will use the parameter $\eta$ instead of the time; the derivatives 
will be calculated with respect to $\eta$ using relation (\ref{2}), so 
that 
\[
\Dr {}{t}=\frac {c}{R(\eta)}\Dr {}{\eta}.
\]
The derivative of the frequency with respect to the time is the 
easiest to find. Due to the redshift,
\begin{eqnarray*}
&\disp \nu=\nu_0\frac {R_0}{R},\quad x=\frac {\nu}{\nu_\st}=\frac
{\nu_0}{\nu_\st}
\frac {R_0}{R}=x_0\frac {R_0}{R}, &\\
&\disp \DR {x}{t}=-x_0R_0\frac {\dot {R}}{R^2}=-xH,&
\end{eqnarray*}
where $\disp H=\frac{\dot{R}}{R}$ is the Hubble function and $\nu_\st$
is some specified frequency.

To determine the other derivatives, we must analyze the variations 
of the corresponding values along the ray.

\subsubsection{ Left-hand Side of the Transfer Equation}
Let us consider the ray along which the radiation propagates, having, 
in general, already been scattered. An arbitrary ray can be described 
using several parameters, one of which can be chosen to be the coordinates 
of the point on the ray closest to a point source at the coordinate origin. 
We will call this the initial point. Due to the symmetry of the problem, 
it is sufficient to specify this distance using the coordinate $\chi_\o$. 
Let a photon pass through the initial point at time $t_\o=t(\eta_\o)$. 
Its position on the ray will be described by the parameter $\chi_*$ 
associated with this time and measured from time $t_\o$. Let us 
determine the derivatives with respect to $\chi_*$  of the variables 
describing the position of the photon on the ray.

\begin {figure}[ht]
\centerline {\psfig {file=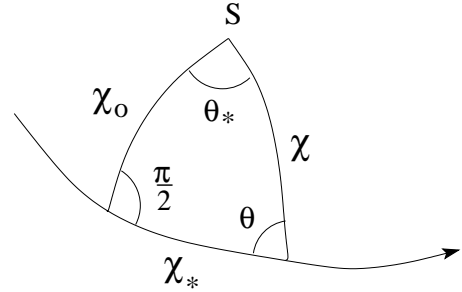,height=4cm}}
\caption {The curvilinear triangle.}
\end {figure}

To this end, let us consider the curvilinear triangle presented in Fig. 1, 
with its sides extending from the source S to the initial point, from 
the source to the photon at time $\eta$, and from the initial point 
to the photon. Since all the lengths are proportional to the radius 
of curvature, we will consider dimensionless distances described by 
the coordinates $\chi_0$, $\chi$, and $\chi_*$, respectively; the 
corresponding dimensionless lengths are $a_\o=\aaa_k(\chi_\o),\,\aaa_k(\chi)$, 
and $\aaa_k(\chi_*)$. In the same order, the angles opposing these 
sides are those between the ray and radial direction, $\theta$, between 
the radial direction toward the initial point and the ray (a right angle), 
and between the radial direction toward the initial point and the direction 
toward the photon's position, $\theta_*$.

We can write the geometrical integrals of the transfer equation for 
the constructed right-angle curvilinear triangle:
\beq
\disp a_\o\!=\!\aaa_k(\chi_\o)\!=\!\aaa_k(\chi)\sin\theta, \label{161} \\
\disp \frac {\aaa_k(\chi_\o)}{\rrr_k(\chi_\o)}\!=\!\aaa_k(\chi_*)\tg\theta,
\label{162}  \\
\disp r_\o=\rrr_k(\chi_\o)\!=\!\frac {\aaa_k(\chi)}{\aaa_k(\chi_*)}\cos\theta\!=\!
\frac {\cos\theta}{\sin\theta_*}, \label{163} 
\eeq

\begin {eqnarray*}
\strut &\rrr_k(\chi)=r_\o\rrr_k(\chi_*),&\\
&\disp\frac {\aaa_k(\chi)}{\rrr_k(\chi)}\cos\theta=\frac
{\aaa_k(\chi_*)}{\rrr_k(\chi_*)},&\\
&\quad \cos\theta_*=\rrr_k(\chi_*)\sin\theta. &
\end {eqnarray*}
The functions $\aaa_k$ and $\rrr_k$ are specified by equalities (\ref{6}) and 
(\ref{7}). Relation (\ref{163}) follows from the two preceding expressions 
(\ref{161}) and (\ref{162}). These relations are trivial in the case of 
a flat space. For a space with positive curvature, the triangle can be 
placed on a unit sphere in three-dimensional space, with the subsequent 
use of spherical geometry. To make the translation to the case of negative 
curvature, the trigonometric functions expressing the sides of the triangle 
must be changed to hyperbolic functions.

The following derivatives can easily be obtained from these relations:
\beq
\DR {\chi}{\chi_*}=\cos\theta,
\label{17:1} \\
\DR {\theta}{\chi_*}=-\sin\theta\frac {\rrr_k(\chi)}{\aaa_k(\chi)},
\label{17:2}  \\
\DR {\theta_*}{\chi_*}=\frac {\sin\theta}{\aaa_k(\chi)}.
\label{17:3}  
\eeq
The equation of motion of a photon along a ray is $\chi_*=\eta-\eta_\o$, 
so that the derivatives (\ref{17:1}), (\ref{17:2}), and (\ref{17:3}) are 
simultaneously the derivatives with respect to the time coordinate $\eta$.

Thus, the transfer equation (\ref{15})  takes the form 
\begin {eqnarray}
&\disp\frac {c}{R(\eta)}\left[\Dr {n}{\eta}+\cos\theta\Dr {n}{\chi}-
\sin\theta\frac {\rrr_k(\chi)}{\aaa_k(\chi)}\Dr {n}{\theta}\right] &\nonumber\\
&\disp -xH\Dr {n}{x}=c\cI_\c.&
\label{18}
\end {eqnarray}
For a flat space, when $k=0,\,\aaa_0(\chi)=\chi$,\, and $\rrr_0(\chi)=1$, the 
equation takes the usual form.

\subsection{ Right-hand Side of the Transfer Equation}
In the problem we are considering, bremsstrahlung absorption and emission, 
as well as Thomson scattering, are important processes in the interaction 
between the matter and point-source radiation.

Bremsstrahlung absorption by thermal nonrelativistic electrons with a 
Maxwellian velocity distribution and the temperature $T$ is described 
by the formula \cite{VVScurs}
\begin{eqnarray*}
\alpha_{\c\c}^*=n_\e n^{+}k_{\c\c}(\nu,T), \\
\alpha_{\c\c}=(1-e^{-h\nu/\kB T})\alpha_{\c\c}^*, 
\end{eqnarray*}
where the cross section is
\begin{eqnarray*}
k_{\c\c}(\nu,T)=\frac {k_{\c\c}^0}{T^{1/2}\nu^3}, \nonumber\\
k_{\c\c}^0\!=\! 8\pi\frac {e^6}{c\hbar}\frac {\kB}{(6\pi m\kB)^{3/2}}\!=\! \nonumber
3.69\!\times\! 10^8\,\!\hbox {cm$^5$ K$^{1/2}$s$^{-3}$}.
\end{eqnarray*}
The absorption coefficient has been corrected for stimulated emission. 
The Gaunt factor is taken to be unity. The bremsstrahlung coefficient in 
the occupation numbers is related to the absorption coefficient by the 
Kirchhoff-Planck relation:
\begin {eqnarray}
\epsilon_{\c\c}=\alpha_{\c\c}e^{-h\nu/\kB T}.
\label{19}
\end {eqnarray}
The absorption coefficient associated with Thomson scattering is 
$\alpha_\T=n_\e\sigma_\T$, where the cross section is
$\disp\sigma_\T=\frac {8\pi}{3}
\left(\frac {e^2}{mc^2}\right)^2=6.66\times 10^{-25}\!\,$ {cm}$^2$.
The indicatrix for Thomson scattering is the Rayleigh index. Since the 
radiation field does not depend on the azimuth, the equation will contain 
the azimuthaveraged indicatrix
\begin {eqnarray*}
p(\theta,\theta')=1+\frac {1}{2}P_2(\cos\theta)P_2(\cos\theta'),
\end{eqnarray*}
where $P_2(\mu)$ is the second Legendre polynomial.

Thus, the right-hand side of the transfer equation has the form
\begin {eqnarray}
\cI_\c=-[\alpha_{\c\c}(\nu,T)+\alpha_\T]n(\eta,\chi,\theta,x)\nonumber \\
+\epsilon_{\c\c}(\nu,T)+\alpha_\T \overline {I}+
\eps_*(\eta,\chi,\theta,x), \label{20}\\
\disp\overline {I}=\frac{1}{2}\intl_0^{\pi}
\sin\theta'p(\theta,\theta')\d\theta'
n(\eta,\chi,\theta',x) \nonumber
\end {eqnarray}
where $\eps_*$  describes the primary radiation from the point source.

The integral in (\ref{20}) can be divided into two terms containing two 
angular moments $n(\eta,\chi,\theta,x)$:
\begin {eqnarray*}
&\disp n_0(\eta,\chi,x)=\frac {1}{2}\intl_0^{\pi}\sin\theta'\d\theta'
n(\eta,\chi,\theta',x),& \\
&\disp \! n_2(\eta,\chi,x)\!=\!\!\frac {5}{2}\!\!\intl_0^{\pi}
\!\!\sin\theta'\!P_2(\cos\theta')\d\theta'n(\eta,\chi,\theta'\!,x).&
\end {eqnarray*}

Let us now specify the source function $\eps_*$ and separate the direct 
from the diffuse radiation.

\subsection{ Direct Radiation}
The entire space is filled by homogeneous and isotropic equilibrium 
cosmological radiation with the mean occupation number
\begin {eqnarray}
\!\!\!n_\e(\eta,x)\!=\!\!\frac {1}{\disp e^{h\nu_\st x/(\kB T)}\!-\!1}\!=\!\!
\frac {1}{\disp e^{h\nu_\st x_0/\kB T_0}\!-\!1}.
\label{21}
\end {eqnarray}
It is straightforward to verify that this function satisfies (\ref{18}) 
with the right-hand side (\ref{20}) without the last term. In fact, both 
the left-hand and right-hand sides vanish, the left-hand side because 
the mean occupation number for the background radiation does not depend 
on anything, and the right-hand side due to the Kirchhoff-Planck relation 
(\ref{19}). This is the equilibrium background radiation. In spite of the 
fact that it interacts with matter, overall, the matter and radiation do 
not affect one another during this interaction; only a common temperature 
is established. After the epoch of recombination, the electron density 
becomes extremely small (it can be assumed to be zero), and the right-hand 
side of the transfer equation no longer contains terms describing the 
interaction between radiation and matter (the radiation is separated 
from the matter), while the left-hand side corresponds to the free 
propagation of the radiation. The quantity (\ref{21}) remains constant 
and satisfies the transfer equation.

Let us now consider the direct radiation of the
source. We will first find an expression for $\eps_*$. The relation 
between the total energy $L$ emitted by the source (for all time, in 
all space, in all directions and frequencies) and  $\eps_*$ is
\begin {eqnarray*}
 L=\frac {2h\nu_\st^4}{c^2}\intl_0^\infty\d t\int\d^3r\intl_0^\infty x^3\d x\,
\eps_*(\eta,\chi,\theta,x).
\end {eqnarray*}
The factor before the integrals is introduced to translate the 
occupation number into intensity. Expressing time in terms of $\eta$ and 
writing the integral over space in terms of integrals over the three 
variables, taking into account the volume element
$\d^3r=R^3(\eta)\aaa_k^2(\chi)\sin\theta\d\chi\d\theta\d\ff$, we obtain
\begin {eqnarray}
&\disp L=2\pi\frac {2h\nu_\st^4}{c^3}\intl_0^\infty
R^4(\eta)\d\eta\intl_0^\infty
\aaa_k^2(\chi)\d\chi & \nonumber\\
&\disp \times\intl_0^\pi\sin\theta\d\theta\intl_0^\infty x^3\d x\,
\eps_*(\eta,\chi,\theta,x). &
\label{22}
\end {eqnarray}
It follows from the expression for the right-hand side of the transfer 
equation (\ref{20}) that the units of $\eps_*$ are inverse length.

If the power of the source radiation at the time $t_\s$ with the time 
coordinate $\eta_\s$ at frequency $x_\s$ is $L\delta(t-t_\s)$, we can write 
\begin {eqnarray*}
&\disp \eps_*(\eta,\chi,\theta,x)=L\frac {c^3}{2h\nu_\st^4}
\frac {\delta(\eta-\eta_\s)}{R^4(\eta_\s)}
\frac {\delta(\chi)}{\aaa_k^2(\chi)} &\nonumber\\
&\disp \times\frac {\delta(\cos\theta-1)}{2\pi}\frac {\delta(x-x_\s)}{x_\s^3},&
\end {eqnarray*}
and relation (\ref{22}) is satisfied.

Further, let us determine the mean occupation number for the direct 
radiation from the source. It is specified by the transfer equation (\ref{18})
with the right-hand side (\ref{20}) without any terms describing scattering:
\begin {eqnarray*}
& \strut\disp \left[\Dr {}{\eta}+\cos\theta\Dr {}{\chi}-\sin\theta
\frac {\rrr_k(\chi)}{\aaa_k(\chi)}\Dr {}{\theta}\right. &\nonumber \\
&\disp \left. -x\frac {a'(\eta)}{a(\eta)}\Dr {}{x}\right]
 n_*(\eta,\chi,\theta,x) &\nonumber \\
&\strut\disp =-[n^{+}k_{\c\c}(\nu,T)+\sigma_\T]n_\e R(\eta)
n_*(\eta,\chi,\theta,x) &\nonumber \\
&\disp +\eps_*(\eta,\chi,\theta,x)R(\eta). &
\end {eqnarray*}
Direct substitution verifies that the solution of this equation is 
given by the function
\begin {eqnarray*}
&\disp n_*(\eta,\chi,\theta,x)=L\frac {c^3}{2h\nu_\st^4}
\frac {\delta(\chi-\eta+\eta_\s)}{R^3(\eta_\s)}
\frac {\Theta(\chi)}{\aaa_k^2(\chi)} & \nonumber \\
&\disp \times\frac {\delta(xa(\eta)/a(\eta_\s)-x_\s)}{x^3_\s}
\frac {\delta(\cos\theta-1)}{2\pi}e^{-\tau}, &
\end {eqnarray*}
where the optical distance from the source to the point is (see below)
\begin {eqnarray*}
&\disp \tau=\tau(\eta,\eta-\eta_\s,x_\s
a(\eta_\s)/a(\eta))=\tau_\T(\eta)-\tau_\T(\eta_\s) &\\
&\disp +\left(1-e^{-h\nu_\st x_\s/\kB T_\s}\right)R_0
\frac {k_{\c\c}^0[E(\eta)-E(\eta_\s)]}{\nu_\st^3x_\s^3T_\s^{1/2}
a^{7/2}(\eta_\s)}.&
\end {eqnarray*}

\subsection{ Discriminating between Direct and Diffuse Radiation}
Let us represent the total occupation number for the photon states 
created by the source as the sum of two terms:
\begin {eqnarray*}
n(\eta,\chi,\theta,x)=n_*(\eta,\chi,\theta,x)+n_\d(\eta,\chi,\theta,x),
\end {eqnarray*}
where $n_\d$ describes the diffuse part of the radiation.
This function satisfies the same transfer equation (\ref{18})
as the initial function, but with the "source" term $\eps_*$
replaced by $\eps_\s(\eta,\chi,\theta,x)$ --- the part of the emission
coefficient due to scattering of the direct radiation from the point 
source. This quantity is specified by the formula
\begin {eqnarray*}
\eps_\s(\eta,\chi,\theta,x)\!=\!\alpha_\T\!\left[1+\frac {1}{2}P_2(\cos\theta)
\right]n_0^*(\eta,\chi,x),
\end {eqnarray*}
where the zeroth moment of the direct intensity is
\begin {eqnarray*}
&\disp n_0^*(\eta,\chi,x)=\frac {L}{4\pi}\frac {c^3}{2h\nu_\st^4}
\frac {\delta(\chi-\eta+\eta_\s)}{R^3(\eta_\s)}
\frac {\Theta(\chi)}{\aaa_k^2(\chi)} &\\
&\disp \times\frac {\delta(xa(\eta)/a(\eta_\s)-x_\s)}{x^3_\s}e^{-\tau}.&
\end {eqnarray*}
The second moment of the direct radiation exceeds the zeroth moment by 
a factor of five, since $P_2(1)=1$.

For a noninstantaneous and nonmonochromatic source, the emitted energy 
may depend on $\eta_\s$ and $x_\s$, and all the equations and their 
solutions must be integrated over these parameters. This does not 
present a problem, due to the presence of $\delta$ functions in 
these expressions.
\subsection{ A Planar Source}
For comparison, a layered planar source can be considered. 
This is only possible if the geometry of the space itself is 
flat, $k=0$. As always in a planar geometry, the equation of 
radiative transfer does not contain an angular derivative, since the 
angle $\theta$ between the direction of propagation of the radiation 
and the normal to the source plane along the ray remains constant. Thus, 
the transfer equations for a planar geometry can be obtained from the 
corresponding equations for spherical geometry, simply by omitting the 
term with the derivative with respect to the angle $\theta$. In the 
expression for the source term and the intensity of the direct radiation, 
the combination $\delta(\chi)/\aaa_k^2(\eta)$ must be replaced by 
$\delta(\chi-\chi_\s)$. 

Thus, the transfer equation acquires the form
\begin {eqnarray*}
\frac {c}{R(\eta)}\left[\Dr {n}{\eta}+\cos\tet\Dr {n}{\chi}\right]-
xH\Dr {n}{x}=c\cI_\c,
\end {eqnarray*}
where the collision integral $\cI_\c$ is specified by the same 
formula (\ref{20}). In this case, the source term is
\begin {eqnarray*}
&\disp \eps_*(\eta,\chi,\mu,x)=L_\s\frac {c^3}{2h\nu_\st^4}
\frac {\delta(\eta-\eta_\s)}{R^2(\eta_\s)}\delta(\chi-\chi_\s) &\\
&\disp \times\frac {\delta(\cos\tet-\cos\tet_\s)}{2\pi}
\frac {\delta(x-x_\s)}{x_\s^3}.&
\end {eqnarray*}
The source is assumed to be located at the level with $\chi_\s$. 
Here, we calculate its luminosity per area of the planar boundary; for 
this reason, the denominator contains a second power of the radius of 
curvature. We assume that the source radiation does not depend on 
azimuth and propagates at an angle $\tet_\s$ relative to a normal to 
the layers.
 
\section{Optical distances}
\subsection{ Optical Distance from the Source}
Under the adopted conditions, the densities of electrons $n_\e$ and 
protons $n^{+}$ depend only on time (and, of course, the cosmological 
model). In the case of total ionization, $n^{+}=n_\e=n_\e^0/a^3(\eta)$; 
in the case of partial ionization, the ionization equation must be 
solved. We will assume that the matter consists of completely ionized hydrogen. 

Expressions for the coefficient of bremsstrahlung absorption by thermal 
nonrelativistic electrons and the absorption coefficient for Thomson 
scattering were given in Section 2.

Since the scattering coefficient does not depend on frequency, the 
corresponding optical distance can be determined very easily. If a 
photon was emitted from the source at time $\eta_\s$, its equation 
of motion is $\chi=\eta-\eta_\s$. In the course of its motion, the 
optical distance between the photon and the source will increase 
according to the relation
\begin {eqnarray*}
&\disp
\tau_\T(\eta,\eta_\s)=\intl_0^\chi\alpha_\T(\eta_\s+\chi')R(\eta_\s+\chi')
\d\chi' & \\
&\disp=\intl_{\eta_\s}^\eta\alpha_\T(\eta')R(\eta')\d\eta'=\tau_\T(\eta)-
\tau_\T(\eta_\s), &
\end {eqnarray*}
where the universal Thomson opacity is
\begin {eqnarray}
&\disp \tau_\T(\eta)=\intl_{\eta_\i}^\eta\alpha_\T(\eta)R(\eta)\d\eta
& \nonumber\\
&\disp =\sigma_\T\intl_{\eta_\i}^\eta n_\e(\eta)R(\eta)\d\eta.&
\label{23}
\end {eqnarray}
Since only differences in the optical distances are of interest, the 
lower integration limit in this formula can be chosen arbitrarily.

The situation with the bremsstrahlung mechanism is more complicated, 
since in this case, the absorption cross section depends on both the 
frequency and temperature. The time dependence of the frequency is 
described as follows. Let the frequency of a photon emitted from the 
source at time $t_\s=t(\eta_\s)$ be $\nu_\s$. Then, its frequency at 
time $t=t(\eta)$, will be $\nu=\nu_\s a(\eta_\s)/a(\eta)$. The 
temperature also depends on the time, according to the usual formula, 
$T\!=\!T_0/a(\eta)\!=\! T_\s a(\eta_\s)/a(\eta)$, where 
$T_\s=T_0/a(\eta_\s)$ is the temperature when the photon is emitted. 
The ratio $\nu/T=\nu_\s/T_\s$ is time independent.

Since the particle densities, frequency, and temperature depend on time 
in a known way, the absorption coefficient is a known function of $\eta$ 
or $z$:
\begin {eqnarray*}
&\disp \strut\disp \tau_{\c\c}(\eta,\eta_\s,x)\!=\!
\left(1\!-\!e^{-h\nu_*x_\s/\kB T_\s}\right)\!\!\intl_0^\chi\!
n_\e(\eta_\s\!+\!\chi') & \\
& \disp \times n^{+}(\eta_\s\!+\!\chi')\frac {k_{\c\c}^0}
{\nu_\st^3 x_\s^3a^{7/2}(\eta_\s)T_\s^{1/2}}a^{7/2}(\eta_\s\!+\!\chi') & \\
&\disp \times R(\eta_\s+\chi')\d\chi'\!
 \strut\disp =\left(1-e^{-h\nu_\st x/\kB T}\right)R_0 &\\
&\disp \times k_{\c\c}(\nu_\st x_\s a(\eta_\s),T_\s a(\eta_\s))[E(\eta)-
E(\eta_\s)], &
\end {eqnarray*}
where
\begin {equation}
E(\eta)=\intl_{\eta_\i}^\eta n_\e(\eta)n^{+}(\eta)a^{9/2}(\eta)\d\eta
\label{24}
\end {equation}
is a quantity similar to the emission measure. The total optical distance 
from the source is
\begin {eqnarray*}
\tau(\eta,\eta_\s,x)=\tau_\T(\eta,\eta_\s)+\tau_{\c\c}(\eta,\eta_\s,x).
\end {eqnarray*}
\subsection{ Calculation of the Scattering Optical Distance}
Let us choose for the initial times in (\ref{23}) and (\ref{24}) the 
times $t_\i=t(\eta_\i)$ for which the ratio $y=\aaa_k(\eta)/\uuu_k(\eta)$ 
assumes the values from the table. The table also presents the products 
$\disp X=\frac {a_\r}{a_\d}y=By$ for the same times. Here, we have chosen 
simple values for this ratio; in fact, the values of $\eta_\i$ are 
maximum for $k=0$ and $k=-1$, $\eta<\eta_\i$, so that the optical distance 
(\ref{23}) will be negative. This does not cause any problems, since only 
the differences of these quantities, which are positive, will be needed. 
When $k=1$, the value of $\eta_i$ is finite, since the range of variations of 
this variable is finite in a closed model.

We will rewrite the integral (\ref{23}), substituting the time dependence 
of the electron density, $n_\e=n_\e^0/a^3(\eta)=(\rho_\b^0/m_\H)/a^3(\eta)$, 
where $\rho_\b^0$ is the current baryon density:
\begin {eqnarray*}
\tau_\T(\eta)=\sigma_\T\intl_{\eta_\i}^\eta n_\e R(\eta)\d\eta=\sigma_\T
\frac {\rho_\b^0}{m_\H}R_0\intl_{\eta_\i}^\eta\frac {\d\eta}{a^2(\eta)}.
\end {eqnarray*}
Introducing the new variable of integration $\disp X=\frac {a_\r}{a_\d}y=By$, 
where $y$ is specified by (\ref{9}), and using the characteristics of this 
ratio enables us to reduce the integral to the form
\begin {eqnarray*}
&\disp \intl_{\eta_\i}^\eta\frac {\d\eta}{a^2(\eta)}=-\frac {1}{2}
\intl_{y_\i}^y\frac {y^2+k}{(a_\r y+a_\d)^2}\d y & \\
&\disp=-\frac {1}{2}\frac {a_\d}{a_\r^3}f_\T(X,B,k), & \\
&\disp f_\T(X,B,k)=\intl_{X_\i}^X\frac {X_1^2+kB^2}{(X_1+1)^2}\d X_1.&
\end {eqnarray*}
Calculating the integral within the limits indicated in	the table 
yields for $k=0$ and 1:
\begin {eqnarray}
\disp \!\! f_\T(X,B,k)\!=\!X\!\frac {X\!+\!2\!+\!kB^2}{X\!+\!1}\!-\!2
\ln(X\!+\!1).
\label{25}
\end {eqnarray}
When $k=-1$, the constant $B(B-2)+2\ln(1+B)$ is added, so that, in an 
open model, this function can also be written in a form that explicitly 
vanishes for the limiting value of the argument, $X=B$:
\begin {eqnarray*}
&\disp f_\T(X,B,-1)=(X-B)\frac {X-B+2}{X+1} & \\
&\disp -2\ln\frac {X+1}{B+1}.&
\end {eqnarray*}
For arguments close to the limiting value, $X\to 0$, the function (\ref{25}) 
is equivalent to $f_\T(X,B,k)\sim kB^2X(1-X)+(1/3+kB^2)X^3$, while for an open 
model, we obtain as $X\to B$ then 
$\disp f_\T(X,B,-1)\sim\frac {(X-B)^2}{(B+1)^2}\left[B+\frac {1/3-B}{B+1}(X-B)
\right]$.
\begin{table}
\caption{$\eta_i$, $y_i$ and $X_i$}
\bigskip
\begin{tabular}{|c|c|c|c|c|c|c|c|}
\hline
$k$ & $y$ & $\eta_\i$ & $y_\i$ & $X_\i$ \\
\hline
$k=1$  & $\ctg(\eta/2)$ & $\pi$    & $0$ & $0$ \\
$k=0$  & $2/\eta$     & $\infty$ & $0$ & $0$ \\
$k=-1$ & $\cth(\eta/2)$ & $\infty$ & $1$ & $a_\r/a_\d$ \\
\hline
\end{tabular}
\end{table}

Thus, the optical distance based on Thomson 
scattering is given by the formula 
\begin {eqnarray}
&\disp\tau_\T(\eta,\eta_\s)=\frac {\sigma_\T}{4}\frac {\rho_\b^0}{m_\H}
\frac {\rho_\d^0}{\rho_\r^0}\left(\frac {8\pi G}{3c^2}\rho_\r^0\right)^{-1/2} &
\nonumber \\
&\disp\times [f_\T(X_\s,B,k)-f_\T(X,B,k)],&
\label{26}
\end {eqnarray}
where $\disp B=\frac {a_\r}{a_\d}$, $\disp X=By=B\frac {\aaa_k(\eta)}
{\uuu_k(\eta)}$, and $X<X_\s$ for $\eta>\eta_\s$.
The factor before the bracket can also be written as
$\disp\frac {\sigma_\T}{4}\frac {\rho_\b^0}{m_\H}
\frac {\Omega_\d^0}{(\Omega_\r^0)^{3/2}}\frac {c}{H_0}$.
\subsection{ Calculation of the Absorption Optical Distance}
Let us calculate the function (\ref{24}). After making the same 
substitutions as for the Thomson-scattering distance calculation, we obtain
\begin {eqnarray*}
& \disp E(\eta)=\left(\frac {\rho_\b^0}{m_\H}\right)^2\intl_{\eta_\i}^\eta
\frac {\d\eta}{a^{3/2}(\eta)} & \\
& \disp =-\left(\frac {\rho_\b^0}{m_\H}\right)^2\frac {1}{\sqrt {2}}
\intl_{y_\i}^y\frac {\sqrt {y^2+k}}{(a_\r y+a_\d)^{3/2}}\d y & \\
& \disp =-\left(\frac {\rho_\b^0}{m_\H}\right)^2\frac {1}{\sqrt {2}}
\frac {a_\d^{1/2}}{a_\r^2}f_\c(X,B,k),&
\end {eqnarray*}
where
\[
f_\c(X,B,k)=\intl _{X_\i}^X\frac {\sqrt {X_1^2+kB^2}}{(X_1+1)^{3/2}}\d X_1.
\]
Only for $B=0$(i.e. for $k=0$) can the integral be 
expressed in terms of elementary functions: 
\begin {eqnarray}
&\disp f_\c(X)\!=\!f_\c(X,0,0)\!=\!2\sqrt {X\!+\!1}\! & \nonumber \\
&\disp +\!\frac {2}{\sqrt {X\!+\!1}}\!-\!4\!
=\!2\frac {(\sqrt {X\!+\!1}\!-\!1)^2}{\sqrt {X\!+\!1}}\! & \nonumber \\
&\disp =\!\frac {2}{\sqrt {X\!+\!1}}\frac {X^2}{(\sqrt {X\!+\!1}\!+\!1)^2}.&
\label{27}
\end {eqnarray}
If $B\ne 0$, the integral can be expressed in terms 
of elliptical integrals, but it is easier to calculate it 
numerically. 
\subsection{Calculations of Optical Distances for the Nonvacuum Model}
For the values in (\ref{26}), we will obtain $B=5.321\times 10^{-2}$.

Let us consider optical distances in the selected 
model. We will first calculate the dimensionless functions
 $f_T$ and $f_c$ for redshifts $z$ from $10^8$ to $10^2$.
Assuming that the photon was emitted from the source 
at time $\eta_{\s}$, the Thomson-scattering distance at time 
$\eta$ is specified by (\ref{26}). The factor before the bracket in 
this formula is equal to $156.2$. 

The optical distance based on absorption can be 
determined in a similar way. After substituting expressions
 for its terms, the product $R_0E(\eta)$ is shown to be
\begin {eqnarray*}
&\disp R_0 E(\eta)=-\frac {1}{2}\left(\frac {\rho_\b^0}{m_\H}\right)^2
\frac {\sqrt {\rho_\d^0}}{\rho_\r^0}\left(\frac {8\pi G}{3c^2}\right)^{-1/2} &
\\
&\disp \times f_\c(X,B,-1)
 =-\frac {1}{2}\left(\frac {\rho_\b^0}{m_\H}\right)^2
\frac {\sqrt {\Omega_\d^0}}{\Omega_\r^0}\frac {c}{H_0} & \\
&\disp \times f_\c(X,B,-1).&
\end {eqnarray*}
 Let us estimate the optical distance for the frequency 
at which the Planck function reaches its maximum 
at a certain cosmological time, $\nu_\m\!=\!c_\W\kB T/h,\,c_\W\!=\!2.821438$.
 For this distance, 
\begin {eqnarray}
&\disp \tau_{\c\c}\!=\!(1-e^{-c_\W})\left(\frac {\rho_\b^0}{m_\H}\right)^2
\!\frac {k_{\c\c}^0}{T_0^{7/2}}\left(\frac {h}{c_\W\kB}\right)^3 \!\!
\frac {\sqrt {\Omega_\d^0}}{\Omega_\r^0} & \nonumber \\
&\disp \times\frac {c}{2H_0}[f_\c(X_\s,B,-1)-f_\c(X,B,-1)].&
\label{28}
\end {eqnarray}
\subsection{Calculation of Optical Distance for a Flat Model} 
The optical distances needed for our model have 
already been determined. Let us rewrite them using 
new notation. The scattering optical distance is 
\begin {eqnarray}
& \strut\disp {\tau_\T=\sigma_\T\frac {\rho_\b^0}{m_\H}R_0
\frac {2c}{\tilde {H}_0R_0}\frac {\Omega_0}{8(1-\Omega_0)^{3/2}}} &\nonumber \\
& \strut\disp {\times\left[f_\T\left(\frac {\sqrt {1-\Omega_0}}{\Omega_0}
\frac {2}{\zeta_\s}\right)
-f_\T\left(\frac {\sqrt {1-\Omega_0}}{\Omega_0}\frac {2}{\zeta}\right)\right]}
& \nonumber \\
& \strut\disp =\frac {\sigma_\T}{4}\frac {\rho_\b^0}{m_\H}\frac {c}
{\tilde {H}_0}\frac {\Omega_0}{(1-\Omega_0)^{3/2}}& \nonumber \\
& \disp\times[f_\T(X_\s)-f_\T(X)]. &
\label{29}
\end {eqnarray}
We have obtained the same formula, (\ref{26}); the 
function $f_T$ coincides with that determined from (\ref{25}), 
but, since $B=0$ and $k=0$, it does not contain any 
parameters: 
\begin {eqnarray*}
f_\T(X)=X\frac {X+2}{X+1}-2\ln(X+1).
\end {eqnarray*}
Taking into account the relations between the Hubble
 constant and critical parameters (\ref{141}), (\ref{142}), 
and (\ref{143}), we will see that the coefficients before the 
brackets in (\ref{29}) and (\ref{26}) also exactly coincide. 

The absorption distance is calculated in exactly 
the same way. This product is equal to 
\begin{eqnarray*}
R_0 E(\eta)=-\frac {1}{2}\left(\frac {\rho_\b^0}{m_\H}\right)^2
\frac {\sqrt {\Omega_0}}{1-\Omega_0}\frac {c}{\tilde {H}_0}f_\c(X),
\end{eqnarray*}
where the function $f_{\c}(X)$ is specified by the previous 
formula (\ref{27}). The resulting formula for the absorption 
distance is 
\begin {eqnarray}
&\disp \tau_{\c\c}=(1-e^{-c_\W})\left(\frac {\rho_\b^0}{m_\H}\right)^2
\frac {k_{\c\c}^0}{T_0^{7/2}}\left(\frac {h}{c_\W\kB}\right)^3 & \nonumber \\
&\disp \times\frac {\sqrt {\Omega_0}}{1-\Omega_0}\frac {c}{2\tilde {H}_0}
[f_\c(X_\s)-f_\c(X)]. &
\label{30}
\end {eqnarray}
Here, the coefficients in (\ref{28}) and (\ref{30}) also coincide. 
\subsection{Comparing the Two Models} 
The optical distances in the two models considered 
differ only in the functions $f$;in the flat model, $B=0$, 
while this parameter is $B\approx 0.05$ in the open model, 
so that the difference is insignificant. Indeed, this 
difference 
\begin {eqnarray*}
& \disp f_\T(X)-f_\T(X,B,-1)=B^2\frac {X}{X+1} & \\
& \disp -B(B-2)-2\ln(B+1) &
\end {eqnarray*}
is small for both small and large $X$. The term that 
is independent of $X$ disappears when the difference 
of the functions is taken. The difference between the 
functions $f_c$ is more difficult to estimate, since these 
functions have different domains: 
\begin {eqnarray*}
& \strut\disp f_\c(X)\!-\!f_\c(X,B,-1)\!=\!
\intl_0^B\!\frac {X_1}{(X_1+1)^{3/2}}\d X_1\\
&\disp +\intl_B^X\!\frac {X_1-\sqrt {X_1^2-B^2}}{(X_1+1)^{3/2}}\d X_1 & \\
& \strut\disp =\frac {2}{\sqrt {1+B}}\frac {B^2}{(1+\sqrt {1+B})^2} & \\
&\disp +B^2\intl_B^X\frac {1}{(X_1+1)^{3/2}}\frac {\d X_1}{X_1+\sqrt
{X_1^2-B^2}}.&
\end {eqnarray*}
The order of the difference is $B^2$. For small $X\sim B$, 
this can be seen directly, while, in the case of larger $X$, 
the contribution of large values of $X_1$ to the integral is 
small. 

We can describe the behavior of our functions for small and large $X$.
When $X\to 0$,  
\begin {eqnarray*}
f_\T(X)=X^3\sum_{n=0}^\infty(-1)^n\frac {n+1}{n+3}X^n, \\
f_\c(X)=X^2\sum_{n=0}^\infty(-1)^n\frac {(2n+1)!}{2^nn!(n+2)}.
\end {eqnarray*}
For large X, 
\begin {eqnarray*}
\!&\!\!\!\disp f_\T(X)\!=\!X\!-\!2\ln X\!+\!1\!-\!\frac
{1}{X}\sum_{n=0}^\infty\frac {(\!-\!1)^n}{X^n}
\frac {n+3}{n+1}, & \\
&\!\!\!\disp f_\c(X)\!=\! 2X^{1/2}\sum_{n=0}^\infty(-1)^{n-1}
\frac {(2n\!-\!3)!}{(2n)!}\frac {2n\!+\!1}{X^n}\!-\!4.&
\end {eqnarray*}
Here, as usual, we assume that $(-3)!=-1,\,(-1)!=1$. 

For the large redshifts of interest for us or, more 
exactly, for $(1-\Omega_0)z\gg 1$, the parameters $\zeta$, $y$, and 
$X$ behave as follows: 
\begin {eqnarray*}
&\disp \zeta\sim\frac {1}{2\sqrt {1-\Omega_0}}\frac {1}{z},&\\
&\disp y=\frac {1}{\zeta}\sim 2\sqrt {1-\Omega_0}z,        & \\
&\disp X=2\frac {\sqrt {1-\Omega_0}}{\Omega_0}y\sim
4\frac {1-\Omega_0}{\Omega_0}z.  &
\end {eqnarray*}
Accordingly, the functions $f$ are easy to determine for 
such $z$ values. The scattering optical distances calculated
 for the open and flat models are only different for 
the largest $z$, and even then only slightly: for $z=10^8$, 
in the fourth significant figure. The absorption distances
 display larger but also insignificant differences, 
in the third significant figure for all $z$. 

Figure 2 presents the results of these calculations. 
The starting value for the curves is $z=z_\s$; $\lg z_\s=n_\s=4(1)8$. 
To present the curves on comparable 
scales, the $\tau_\T$ values on the curve with $\lg z_\s=n_\s$ 
have been multiplied by $10^{1-n_\s}$, while the $\tau_{\c\c}$ curves 
have been multiplied by $10^{(17-n_\s)/2}$.The difference 
between these factors reflects the fact that the absorption
 coefficient is proportional to $T^{-1/2}$. 
\begin{figure}[h]
\begin{center}
\begin{picture}(250,100)
\epsfxsize=8cm \epsfbox{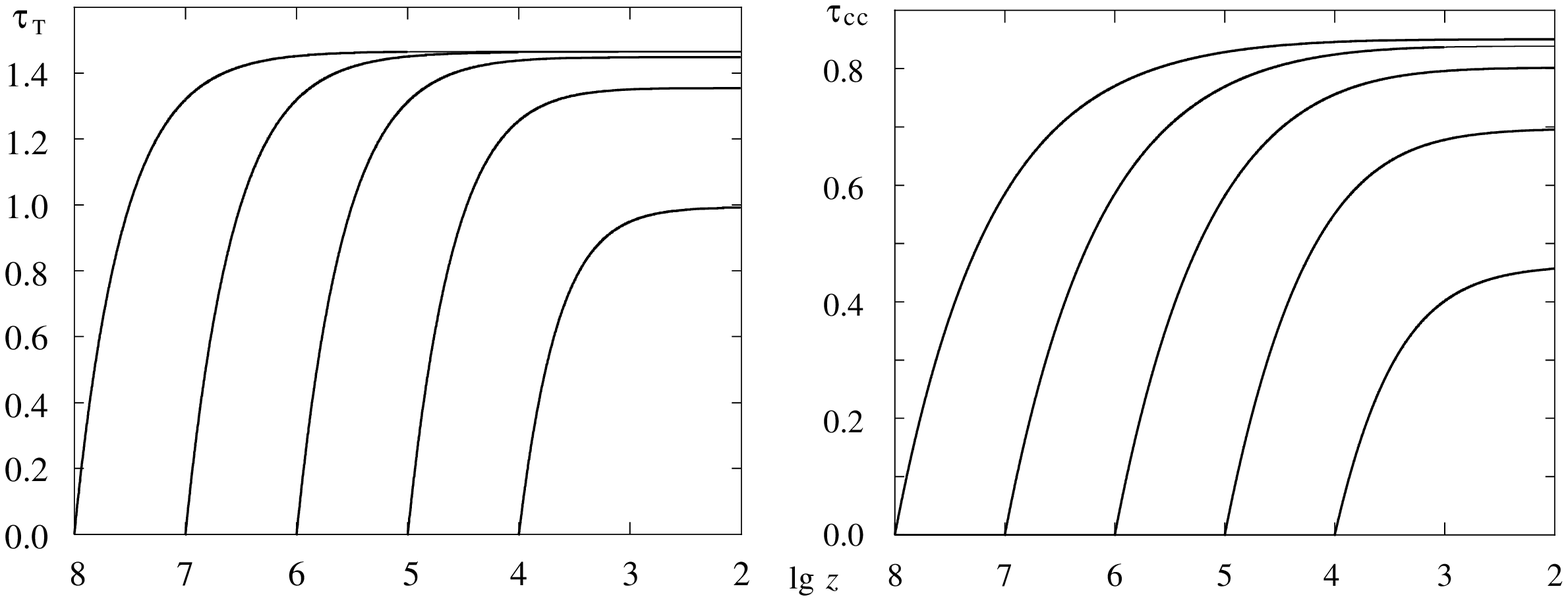}
\end{picture} 
\caption{ Scattering and absorption optical distances as
a function of the photon redshift in the course of the 
photon's motion away from the source. The $\tau_\T$ values are 
multiplied by $10/z_\s$, and the $\tau_{\c\c}$ values by $10^{-8.5}z_\s^{-1/2}$.}
\end{center}
\end{figure}

We can see from the figures that the optical distances
 initially increase rapidly with decreasing redshift;
 this growth is slower for the absorption distances
 at the frequency of the maximum of the Planck 
function. If the source's redshift is $z=z_\s$, then, for 
$z_\s/3\leq z\leq z_\s$, these distances are 
\[
\tau_\T\sim\sigma_\T\frac {\rho_\b^0}{m_\H}\frac {c}{\tilde {H}_0}
\frac {z_\s-z}{\sqrt {1-\Omega_0}},
\]
\[
\tau_{\c\c}\sim (1-e^{-c_\W})\left(\frac {\rho_\b^0}{m_\H}\right)^2
\frac {k_{\c\c}^0}{T_0^{7/2}}
\]
\[
\times\left(\frac {h}{c_\W\kB}\right)^3
\frac {c}{\tilde {H}_0}\frac {\sqrt {z_\s}}{\sqrt {1-\Omega_0}}
\left(1-\frac {z}{z_\s}\right).
\]
For the adopted model parameters, $\tau_\T\!\approx\! 0.13(z_\s\!-\!z)$, and
$\tau_{\c\c}\approx 1.8\times 10^{-9}z_\s^{-1/2}(z_\s-z)$.
The optical distances increase more slowly with decreasing $z$, 
gradually reaching their asymptotic values $\tau_\T\sim 0.14 z_\s$  
 and $\tau_{\c\c}\sim 2.5\times 10^{-9}z_\s^{1/2}$. The distances $\tau_{\c\c}$ 
reach their limiting values at smaller $z$, and these 
values are substantially smaller (by a factor of $5.6\times 10^6 z_\s^{1/2}$)
 than those for the scattering distances. 
Since the coefficient $k_{\c\c}$ is inversely proportional to 
the third power of the frequency, this difference will be 
smaller at lower frequencies. 
\subsection{The Grey Approximation}
Since the scattering coefficient substantially exceeds
 the absorption coefficient, the latter can be neglected
 relative to the source radiation. Equation (\ref{18})
with the right-hand side (\ref{20}) can then be integrated 
over frequency, as is done in calculations of model 
atmospheres \cite{VVScurs}. 

Let us denote the frequency moments for the mean 
occupation number ($s\geq 0$ is not necessarily integer, 
so that this is essentially the Mellin transform): 
\begin {eqnarray*}
n^{(s)}(\eta,\chi,\mu)=\intl_0^\infty x^s n(\eta,\chi,\mu,x)\d x.
\end {eqnarray*}
Integration over the frequency with the weights $x^s$ yields
\begin {eqnarray*}
&\disp \Dr {n^{(s)}}{\eta}+\mu\Dr {n^{(s)}}{\chi}+(1-\mu^2)
\frac {\rrr_k(\chi)}{\aaa_k(\chi)}\Dr {n^{(s)}}{\mu} & \\
&\disp +(s+1)\frac {a'(\eta)}{a(\eta)}n^{(s)}=R(\eta)\cI_\c^{(s)},&
\end {eqnarray*}
where $\cI_\c^{(s)}$ is the result of integrating the right-hand 
side. The integral of the term with the frequency 
derivative was calculated via integration by parts. 
The frequency moments of the collision 
integral are 
\begin {eqnarray*}
&\disp \cI_\c^{(s)}=-\alpha_\T\left[n^{(s)}(\eta,\chi,\mu)-n_0^{(s)}(\eta,\chi)
\right.& \\
&\disp \left.
-\frac {1}{10}P_2(\mu)n_2^{(s)}(\eta,\chi)\right]+\eps_*^{(s)}(\eta,\chi,\mu),&
\end {eqnarray*}
where the moments of the source power are 
\begin {eqnarray*}
\eps_*^{(s)}(\eta,\chi,\mu)=\intl_0^\infty\eps_*(\eta,\chi,\mu,x)x^s\d x.
\end {eqnarray*}
After separating out the direct radiation, the function $\eps_*$
 should be substituted by $\eps_\s$.
\subsection{ CONCLUSION }
We have derived an equation describing the evolution
 of the radiation of a source whose intensity 
exceeds that of the surrounding equilibrium background
 radiation at the epoch of radiation-dominated 
plasma, between the epochs of annihilation and recombination.
 A cosmological model for this period 
was adopted. We have separated the diffuse and direct 
radiation, and calculated the opacities of the matter
 due to Thomson scattering and bremsstrahlung 
absorption. The latter mechanism can be neglected 
relative to the source radiation, and this fact was used 
to obtain an equation for the frequency moments. 

This is the first paper in the series of studies that 
will be concerned with elucidating how sources of 
radiation affect the evolution of the Universe and how 
their presence is reflected in the thermal background 
radiation, taking into account the fact that the accuracy
 of observations is growing steadily. We hope to 
solve this problem in our forthcoming studies. 

\end{document}